\documentclass{ptephy}%%%%where ptephy is the template name
\usepackage{amsmath} 
\usepackage{amsthm} 
\usepackage{cite} 
\usepackage{graphicx} 
\usepackage{algorithmic} 
\usepackage{subfig} 
\usepackage{url} 
\usepackage{latexsym}
\usepackage{natbib}
\usepackage{bm}
\bibliography{sample}
\begin{document}
\title{ Super-horizon second-order perturbations for cosmological random
 fluctuations and the Hubble-constant problem}
\author{\name{Kenji Tomita}{\ast}}
\address{\affil{}{Yukawa Institute for Theoretical Physics, Kyoto University, Kyoto 606-8502,
 Japan}
\rm{\email{ketomita@ybb.ne.jp}}}

\begin{abstract}
The super-horizon second-order density perturbations corresponding 
to cosmological random fluctuations are considered, their non-vanishing
 spatial average is shown to be useful in solving the serious problem 
with the cosmological tension between measured Hubble constants at 
present and those at the early stage, and the difference from 
previous works on the backreaction is discussed.
\end{abstract}

\maketitle

%ch1 ----------------------------------------------------------
\section{Introduction}
There are two types of gravitational instability theories in the expanding
universe. One of them is general-relativistic theory. The linear theory
was first derived by Lifshitz\citep{lif, lk} and the gauge-invariant version 
 by Bardeen\citep{bar} and others\citep{kodama}.  
The second-order nonlinear theory
 in the matter-dominant stage was first derived by us\citep{tomold, tom}
  extending the Lifshitz theory, and by Russ et al. \citep{russ} and 
Matarrese et al. \citep{mat} in  different formulations.
Furthermore their gauge-invariant theories and nonlinear theories 
at the early stages also have been studied thereafter.\citep{bru,naka}
The backreaction problems were studied by Nambu\citep{nb1,nb2} 
analyzing the second-order perturbations that include the renormalization 
process. Kasai et al.\citep{kaf} derived a no-go theorem for an 
accelerating universe through the backreaction. 

Another representative theory is the Newtonian instability theory, which
has been studied by many workers \citep{mak, bern}, in which the 
Newtonian gravitational
potential is treated only linearly, but the hydrodynamical quantities
can be treated in the second order and higher orders. This theory can be
applied only to perturbations whose linear size $L$ is smaller than
$1/H$, where $H$ is the Hubble parameter.  

In this letter, we discuss using these theories on what scales 
the second-order perturbations
for cosmological random fluctuations can have a non-vanishing 
average to solve the Hubble-constant problem, and on the difference
from  previous works. 

For later reference we show here the condition that the inequality 
$L < 1/H$ holds 
always  at the matter-dominant stage after an epoch such as $1 + z_1
= 1500$. In a flat model with the density parameter $\Omega_M$ and
the cosmological-constant parameter $\Omega_\Lambda 
(= 1 - \Omega_M)$, we obtain for the 
present length $L_0 \ (= (1 + z_1) L_1)$       
\begin{equation}
  \label{eq:1}
L_0 < L_{0m}  \equiv 200 h^{-1} /{(15\Omega_M)}^{1/2} 
\end{equation}
from the above condition,
where the present Hubble constant $H_0$ is $100 h$   { km \ s$^{-1}$
 { Mpc}$^{-1}$.    For $\Omega_M = 0.22$ and $0.24$, we have
\begin{equation}
  \label{eq:2}
L_{0m}  = 110 h^{-1} \ {\rm and } \ 105 h^{-1}  \ {\rm Mpc},
\end{equation}
respectively.

In Sect. 2, we show the derivation of second-order perturbations from 
random fluctuations. 
In Sect. 3, we explain the meaning of the renormalized model parameters 
used in our previous papers \citep{tompre, tompre2}, and show their 
influence of the parameters upon the 
 Hubble-constant problem. In Sect. 4, we have concluding remarks
which include the comparison with the previous works on the 
backreaction.

%ch2 ----------------------------------------------------------
\section{Random fluctuations and second-order perturbations}
Now let us consider the random fluctuations that were caused by
quantum fluctuations at the very early stage, and whose amplitude 
and spectrum have been studied through the precise measurements 
of fluctuations in the cosmic microwave background  (CMB) radiation
by WMAP\citep{wmap} and Planck Collaborations\citep{planck1, planck2}.
Here we assume that the fluctuations as  first-order perturbations
$\delta_1 \rho/\rho$ have a vanishing spatial average (i.e. $\langle
\delta_1 \rho/\rho \rangle = 0$), where $\rho$ is the background matter
density, and derive the corresponding second-order perturbation 
$\delta_2 \rho/\rho$ and its spatial average $\langle \delta_2 \rho/\rho \rangle$. 

For perturbations with the present size satisfying the condition 
$L_0 < L_{0m}$, the
they can be treated in the Newtonian cosmological 
approximation\citep{mak, bern}, and so 
\begin{equation}
  \label{eq:3}
\langle \delta_2 \rho/\rho \rangle = 0 
\end{equation}
at the matter-dominant stage (with $1 + z < 1500$). Similarly higher-order
perturbations also have vanishing average values, i.e.
\begin{equation}
  \label{eq4}
\langle \delta_n \rho/\rho \rangle = 0 
\end{equation}
for $n > 2$.

   For the perturbations with $L_0 > L_{0m}$, on the other hand,  the
linear sizes of the perturbations are always larger than the Hubble size
 $1/H$ or cross it once at the matter-dominant stage,  and therefore
  the evolution should be treated
using  general-relativistic perturbation theories.

In one of our recent papers (I) \citep{tompre}, we used the second-order 
perturbations $\delta_2 \rho/\rho$, which were derived using our 
general-relativistic perturbation theory with non-zero $\Lambda$ (in the
 comoving and synchronous
 gauge)  \citep{tom} and include non-Newtonian terms, and obtained the
  spatial average $\langle \delta_2 \rho/\rho \rangle$ in the form
of an integral with respect to wave-number $k$.
In this paper, the spectrum of first-order density
fluctuations was given by using the BBKS transfer function\citep{bbks},
 and the
 amplitude was determined using the result of the Planck measurements.

 The upper limit of the wave-number  $k_{max}$ in the above integral  was
  specified as
\begin{equation}
  \label{eq5}
L_{max} \equiv 2\pi/k_{max} = 102/h  \quad {\rm Mpc}  
\end{equation}
for $\Omega_M = 0.22$.   This length of $L_{max}$ represents
the present distance, over which smooth observations 
on cosmological scales may be possible. The condition that the 
wave-number $k$ should be smaller than $k_{max}$ (or the length $L_0$ of
perturbations should be larger than $L_{max}$)  is nearly equal to 
the super-horizon condition $L_0 > L_{0m}$ with Eq.(2), because 
$L_{max} \simeq L_{0m}$.
Thus, in (I), we obtained $\langle \delta_2 \rho/\rho \rangle$ with
a positive value from the super-horizon random fluctuations with
the length $L_0 > L_{0m}$. This  $\langle \delta_2 \rho \rangle$
is the average energy density of random fluctuations which we call 
the density of the fluctuation energy.

%ch3------------------------------------------------------------------
\section{Energy density of random fluctuations and the Hubble-constant
 problem}
     
In paper (I) \citep{tompre}, we derived the second-order perturbations
 corresponding to not 
only density perturbations but also metric perturbations, and obtained the 
average values of   $\langle \delta_2 \rho\rangle$  and 
$\langle \delta_2 (H^2) \rangle$, where $H$ is the Hubble parameter. 
 
Using these average values, we defined the cosmologically renormalized quantities
\begin{equation}
  \label{eq6}
\rho_{rem} \equiv \rho +  \langle \delta_2 \rho \rangle
\end{equation}
and
\begin{equation}
  \label{eq7}
H_{rem} \equiv  [H^2 +  \langle \delta_2 (H^2) \rangle ]^{1/2},
\end{equation}
which were derived in the comoving and synchronous gauge.
Moreover, the renormalized model parameters were defined as
\begin{equation}
  \label{eq:8}
({\Omega}_M)_{rem} \equiv \Omega_M \frac{1 +  \langle  \mathop{\delta}_2
 \rho/\rho  \rangle}{1 + \langle  \mathop{\delta}_2 \rho/\tilde{\rho} 
  \rangle}
\end{equation}
and
\begin{equation}
  \label{eq:9} 
({\Omega}_\Lambda)_{rem}  \equiv \Omega_\Lambda  \frac{1}{1 + \langle 
 \mathop{\delta}_2
 \rho/\tilde{\rho}  \rangle},
\end{equation}
where $\tilde{\rho} \equiv \rho + \Lambda$.
Here and in the previous papers \citep{tompre, tompre2}, the 
``renormalized'' model parameters
 mean the new model parameters, which are given to  homogeneous models 
with the average second-order perturbations. 
This ``renormalization'' used here is not connected with any dynamical
 renormalization processes, including the gauge-invariant property.
  
At present we face the cosmological tension between the direct 
measurements of the Hubble constant at present epoch and the 
Hubble constant derived from the Planck measurements of the 
CMB anisotropies \citep{h1,h2,h21,h3,h4}.
We showed in the previous paper (I) the possibility of solving the problem on the above
tension, using the renormalized Hubble constant (7) and the renormalized
 model parameters (8) and (9).  For the background model with 
 the present parameters 
$\Omega_M = 0.22, \Omega_\Lambda = 1 - \Omega_M$, and $H_0 = 
67.3 \ {\rm km \ s^{-1} Mpc^{-1}}$,  it was found that we obtain 
$(\Omega_M)_{rem} = 0.305, \ (\Omega_\Lambda)_{rem} = 1 -
(\Omega_M)_{rem}$, and $H_{rem} = 74.0 \ {\rm km \ s^{-1} Mpc^{-1}}$
at the present epoch. On the other hand, the models at the early stage 
($z \gg 1$) have values consistent with  the background ones.
These model parameters and Hubble constants are found 
to represent the observed ones.

In our next paper (II) \citep{tompre2}, we expressed the fluctuation 
energy $\rho_f \
 (\equiv   \langle \delta_2 \rho \rangle)$  as a function of $\rho$ 
 in a  background model, and regarded $\rho_f$ as the density of
 a new constituent pressureless matter. 
 Moreover,  assuming that the total matter density $\rho_T$ is given by
\begin{equation}
  \label{eq:10} 
\rho_T = \rho + \rho_f (\rho),
\end{equation}
we derived a new cosmological model with
\begin{equation}
  \label{eq:8}
ds^2 = g_{\mu \nu} dx^\mu dy^\nu = a^2 (\eta) [-d\eta^2 + \delta_{ij} 
dx^i dx^j ],
\end{equation}
where the Greek and Roman letters denote $0, 1, 2, 3$ and $1, 2, 3$,
 respectively. The conformal time $\eta (= x^0)$ is related to the cosmic
time $t$ by $dt = a(\eta) d\eta$, and  $a (\eta)$ is the new scale factor 
in the universe including the fluctuation energy $\rho_f$.
    
 Then  the velocity vector and energy-momentum tensor of 
pressureless matter are expressed in comoving coordinates as
\begin{equation}
  \label{eq:9}
u^0 = 1/a,  \quad u^i = 0
\end{equation}
and
\begin{equation}
  \label{eq:10}
T^0_0 = -\rho_T, \quad T^0_i = 0, \quad T^i_j = 0.
\end{equation}

    From the Einstein equations, we obtain    
\begin{equation}
  \label{eq:11}
 \rho_T a^2 = 3(a'/a)^2 - \Lambda a^2,
\end{equation}
and the energy-momentum conservation ($T^{\mu\nu}_{;\nu} = 0$) 
gives the relation 
\begin{equation}
  \label{eq:12}
\rho_T a^3 = \rho_T (t_0),
\end{equation}
where $a = 1$ at the present epoch ($t = t_0$) and a prime denotes 
$\partial/\partial \eta$.  
    
For comparison, the usual  background model without the
fluctuation energy is denoted using the suffix $b$,  so that the scale
factor, the Hubble parameter and the density are denoted as
$a^b, H^b$ and $\rho^b$, where $a^b = 1$ at the present epoch.  
Then for the present parameters  $\Lambda_M^b = 0.22, 
\Omega_\Lambda^b = 0.78 $, and $H_0^b = 67.3 \ {\rm km \ s^{-1} 
Mpc^{-1}}$, it is found that the present value of $\beta (\rho^b) 
\equiv \rho_f (\rho^b) /\rho^b$ is $0.552$.

By comparing the solutions of $a$ and $a^b$, we could express 
$\Omega_M, \Omega_\Lambda$ and $H$ as the functions of
$z, \Omega_M^b, \Omega_\Lambda^b$ and $H^b$.
It is found that  the correspondence between $(\Omega_M, H)$ and           
$(\Omega_M^b, H^b)$ depends on the value of the ratio of present matter
 densities $(\rho/\rho^b)_0$. 

    For example, for  $(\rho/\rho^b)_0 = 1.181$, we obtain the
present parameters $\Omega_M = 0.341$ and $H_0 = 73.2 \ {\rm km \
 s^{-1} Mpc^{-1}}$, while for $z \gg 1$, we have $(\Omega_M, H) =
 (\Omega_M^b, H^b)$. Their values for the other $(\rho/\rho^b)_0$
 are shown in the paper (II) \citep{tompre2}. These changes 
 in the Hubble constant and 
 cosmological  parameters may explain the observational difference
 between the direct measurements and the Planck CMB measurements.
  They reflect the situation that the
 universe evolves from the background one to the present one
 with larger Hubble constants,
 corresponding to the increase of the fluctuation energy $\rho_f$.

%ch4------------------------------------------------------------------
\section{Concluding remarks}
Using  our second-order perturbations in the comoving and synchronous
 gauge in the super-horizon region, 
it was shown that the Hubble-constant problem may be solved by 
introducing the renormalized parameters or  building
cosmological models that include the fluctuation energy as 
one of the constituent matters.

This conclusion is different from that of Nambu's works\citep{nb1,nb2} 
which treated the backreaction problem. In them it is found that no 
large-scale motion can arise from cosmological random fluctuations.
The difference between two works comes from
the treatment of the perturbations: Nambu analyzed the backreaction
using the theoretical  renormalization process, while we used the above
 simple model (with  second-order perturbations in the comoving 
 and synchronous gauge) without considering any renormalization process 
with gauge invariant property into account.   
 So we need to have comparative studies of these two treatments.
 
 Our work is consistent with Kasai et al.'s no-go theorem for an
 accelerating universe,\citep{kaf} though they used the Newtonian gauge different 
 from our gauge,  where $\rho_T (\equiv \rho + \rho_f)$ in our paper (II)
 corresponds to $\bar{\rho}$ in their paper and their main equation 
 corresponds to our Eq. (\ref{eq:11}). Our model is decelerating in the 
 zero-$\Lambda$ case, though it has comparatively large Hubble parameters.

 \bigskip
%ack ---------------------------------------------------------
\section*{Acknowledgements}
The author thanks  K. T. Inoue  for many  discussions. 
\bigskip

%ref
%--------------------------------------------------------------------

\end{document}